\documentclass[twocolumn,aps,floatfix,citeautoscript,superscriptaddress,prb]{revtex4-2}
\usepackage{newtxtext}
\usepackage[smallerops]{newtxmath}
\usepackage{amsfonts,amsmath,latexsym,bm,wasysym}
\usepackage{graphicx,xcolor,blindtext,float}
\usepackage[colorlinks,bookmarks=true,citecolor=blue,linkcolor=blue,urlcolor=blue, breaklinks=true]{hyperref}
\usepackage[utf8]{inputenc}
\usepackage{t1enc}

\let\lambda\lambdaup
\newcommand{\ztwo}{\mathbb{Z}_2}

\begin{document} 

\title{Perfectly hidden order and \texorpdfstring{$\mathbb Z_2$}{text} confinement transition in a fully packed monopole liquid}

\author{Attila Szabó}
\email[]{attila.szabo@physik.uzh.ch}
\affiliation{Max-Planck-Institut für Physik komplexer Systeme, Nöthnitzer Str.\ 38, 01187 Dresden, Germany}
\affiliation{Physik-Institut, Universität Zürich, Winterthurerstr.\ 190, 8057 Zürich, Switzerland}

\author{Santiago A. Grigera}
\email[]{sgrigera@fisica.unlp.edu.ar}
\affiliation{{Instituto de F\'{\i}sica de L\'{\i}quidos y Sistemas Biol\'ogicos (IFLYSIB), UNLP-CONICET, La Plata, Argentina}}
\affiliation{Departamento de F\'{\i}sica, Facultad de Ciencias Exactas, Universidad Nacional de La Plata, La Plata, Argentina}

\author{P. C. W. Holdsworth}
\email[]{peter.holdsworth@ens-lyon.fr}
\affiliation{ENS de Lyon, CNRS, Laboratoire de Physique, F-69342 Lyon, France}
\affiliation{French American Center for Theoretical Science, CNRS, KITP, Santa Barbara, CA 93106-4030, USA}

\author{Ludovic D. C. Jaubert}
\email[]{ludovic.jaubert@cnrs.fr}
\affiliation{CNRS, Universit\'e de Bordeaux, LOMA, UMR 5798, 33400 Talence, France}

\author{Roderich Moessner}
\email[]{moessner@pks.mpg.de}
\affiliation{Max-Planck-Institut für Physik komplexer Systeme, Nöthnitzer Str.\ 38, 01187 Dresden, Germany}

\author{Demian G. Slobinsky}
\affiliation{{Instituto de F\'{\i}sica de L\'{\i}quidos y Sistemas Biol\'ogicos (IFLYSIB), UNLP-CONICET, La Plata, Argentina}}

\author{Mauricio Sturla}
\email[]{sturla@fisica.unlp.edu.ar}
\affiliation{{Instituto de F\'{\i}sica de L\'{\i}quidos y Sistemas Biol\'ogicos (IFLYSIB), UNLP-CONICET, La Plata, Argentina}}
\affiliation{Departamento de F\'{\i}sica, Facultad de Ciencias Exactas, Universidad Nacional de La Plata, La Plata, Argentina}

\author{Rodolfo A. Borzi}
\email[]{borzi@fisica.unlp.edu.ar}
\affiliation{{Instituto de F\'{\i}sica de L\'{\i}quidos y Sistemas Biol\'ogicos (IFLYSIB), UNLP-CONICET, La Plata, Argentina}}
\affiliation{Departamento de F\'{\i}sica, Facultad de Ciencias Exactas, Universidad Nacional de La Plata, La Plata, Argentina}

\date{\today}

\begin{abstract} We investigate a variant of spin ice whose degenerate ground states are densely packed monopole configurations. An applied field drives this model through a $\ztwo$ confinement transition. This phase change is a variant of the U(1) Kasteleyn transition, but instead of saturated order the system has fluctuations in the confined phase, and shows critical scaling on both sides of the transition. Remarkably, the magnetic response scales with the critical exponent of the \textit{specific heat} in the 3D Ising universality class. We prove this universality using a Kramers--Wannier duality to map to an Ising model. The dual order parameter maps back to a non-local string order parameter in the original model, invisible to any local probe. 
Further, we describe the transition in terms of a bosonic field theory including a pairing term. 
\end{abstract}

\maketitle

\begin{figure}[b]
\includegraphics[width=\columnwidth]{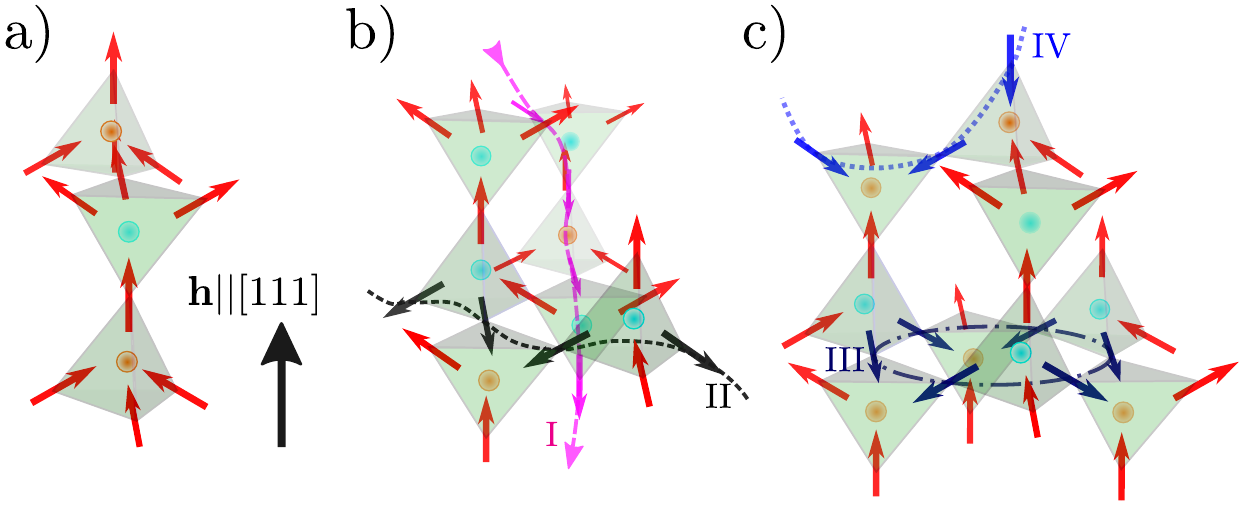}
\caption{ (a)~Ising pyrochlore lattice, saturated in a $[111]$ magnetic field $\mathbf{h}$; single positive (orange) and negative (cyan) charges form a monopole crystal. (b)~On decreasing $h/T$, the system demagnetizes retaining a single charge in each tetrahedron. This may happen through string excitations spanning the whole system in any direction: strings I (in magenta, preserving local charge) and II (in black, inverting sign of the charges it goes through) are respectively along and orthogonal to the [111] field direction.
(c)~Alternatively, finite loop excitations cost a finite energy and may appear at any temperature (III and IV).
Along the [111] axis, the pyrochlore lattice forms alternating kagome and triangular layers.
}
\label{fig:latt}
\end{figure}

\section{Introduction}

Frustrated magnetism has proven a fertile testing ground for the realisation of emergent many-body phenomena as yet  unobserved in high-energy physics \cite{knolle_sl_2019}. The fine structure constant of quantum electrodynamics can be tuned \cite{pace21a}, electromagnetism can become tensorial \cite{pretko20a}, magnetic monopoles and Majorana fermions can be stabilized \cite{castelnovo_2008,spinicebook,Kitaev06a,Hermanns18a}. This is because spin-liquid ground states can often be described by a gauge field, offering a variety of vacua for unusual quasiparticles bearing a gauge charge. Being excitations, these quasiparticles are usually studied at low density, since increasing the temperature populates excited states at the cost of transforming the system into a trivial paramagnet.

Here, we consider the opposite case, analyzing a three-dimensional spin-ice model with maximal monopole density~\cite{slobinsky2018,slobinsky2021,nutakki23b}; in other words, a fully packed monopole (FPM) liquid with a single-charge monopole at every lattice site (see Fig.~\ref{fig:latt}).  It has been shown that the presence of noninteracting monopoles leads to a novel, constrained yet extensively degenerate set of magnetic configurations without introducing spin correlations~\cite{slobinsky2018}. Within this FPM liquid -- contrary to the Coulomb phase in spin ices~\cite{isakov_2004,henley_2010,castelnovo_spinice_2012} -- \textit{every} closed loop of spins can be flipped without escaping the manifold. 

We show both numerically and analytically through a Kramers-Wannier duality \textit{à la} Wegner~\cite{Wegner1971}
that the FPM fluid
has a topological phase transition as a function of field $\mathbf{h}~||$ [111], a rare example of $\ztwo$ (de)confinement in three dimensions (3D). We explain this from the coexistence of two types of excitations:
(i) system-spanning strings of inverted  spins relative to the field direction, similar to those of the celebrated U(1) Kasteleyn transition \cite{kasteleyn63a,bhattacharjee1983,Jaubert2008Kasteleyn}, but running in \textit{all} directions [Fig.~\ref{fig:latt}]; 
and (ii) finite-size loops, present on both sides of the transition~\cite{Nagle1989,smerald16a,smerald18a}. Remarkably, while this transition is driven by a magnetic field, the magnetization (and indeed any other local candidate order parameter) is smooth at the transition, making this an extreme example of hidden order~\cite{chalker1992hidden,mydosh2011colloquium,paddison2015hidden,taillefumier2017competing}.

For more than 50 years, Kasteleyn's mechanism for phase changes~\cite{kasteleyn63a} has independently coexisted with Wegner's work on dualities in Ising models where transitions take place without a local order parameter~\cite{Wegner1971}. The FPM liquid presented here provides a link between these two worlds that have been running parallel to each other until now.

\textit{Model.---}%
We consider Ising spins $\mathbf{S}_i=\sigma_i \mathbf{e}_{\alpha(i)}$ on the pyrochlore lattice, where the $\mathbf{e}_\alpha$ point to the four local $\langle111\rangle$ easy-axis directions and $\sigma_i=+1$ ($-1$) for spins pointing out of (into) ``up'' tetrahedra (see ~\ref{fig:latt}). The Hamiltonian is
\begin{equation}
\mathcal{H} = K\sum_{t}  \sigma_1\sigma_2\sigma_3\sigma_4 - \sum_i \mathbf{h}\cdot\mathbf{S}_i,
\label{eq:ham}
\end{equation}
where $t$ runs over all tetrahedra of the system, $\sigma_{1,\dots,4}$ are the four spins within each tetrahedron, and $\mathbf{h}=h\mathbf{e}_{0}$ is the external magnetic field in reduced units. We work in the $K\to+\infty$ limit, where every tetrahedron is in a ``3-in-1-out'' or ``3-out-1-in'' configuration, so that each tetrahedron hosts a positive or a negative monopole. This defines the FPM model, which has imposed constraints but zero internal energy so that for fixed field direction it is parametrized by a single thermodynamic variable $x \equiv h/T$.  
The field acts as a staggered chemical potential for the monopoles~\cite{castelnovo_2008,raban2019multiple} and orders the emergent gauge fragment of the spins \cite{brooks_artificial_2014}, distinguishing the sublattice $\alpha=0$, whose spins are parallel to $\mathbf{h}$, from the other three, where $\mathbf{h}\cdot\mathbf{e}_\alpha=-h/3$; a large $x$ thus polarizes all spins to a unique monopole-crystal~\cite{Pearce2022} ground state  [Fig.~\ref{fig:latt}(a)].
Along the field direction, the pyrochlore lattice can be seen as an alternating stack of triangular ($\alpha=0$) and kagome ($\alpha=1,2,3$) layers.

\section{Simulations}%
\begin{figure}
    \includegraphics[width=\columnwidth]{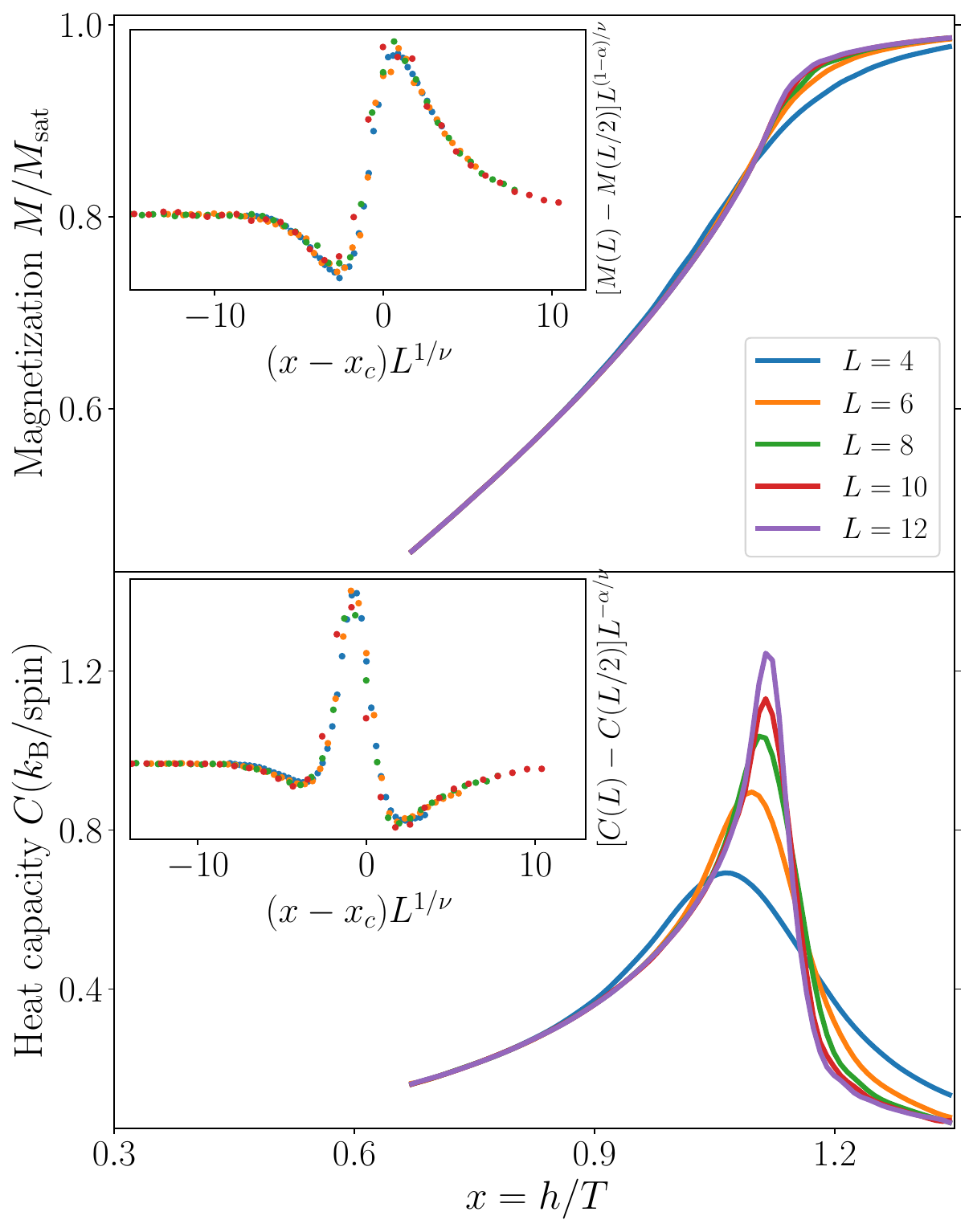}
    \caption{
    Magnetization $M$ (top panel) and specific heat $C$ (bottom panel) as a function of $x=h/T$ for the FPM model in a $[111]$ field. 
    The heat capacity shows a diverging peak
    which obeys a finite-size scaling form with the critical exponents $\alpha,\nu$ of the 3D Ising universality class (inset).
    The magnetic susceptibility is proportional to the heat capacity, so it diverges with the same exponents;
    therefore, the finite-size anomaly in the magnetisation scales with critical exponent $1-\alpha$  (inset). Note $\alpha/\nu=2y_t-d$ and $\nu=1/y_t$ depend uniquely on the thermal eigenvalue exponent $y_t$ consistently with there being a single independent exponent.
    To eliminate non-singular contributions, finite-size scaling was performed on the difference between measurements at system sizes $L$ and $L/2$ (see Appendix \ref{sec: MC}).
    }
    \label{fig:MCv}
\end{figure}

We simulated a system of rhombohedral shape with $N=4L^3$ spins and periodic boundary conditions within each kagome layer, and developed a loop Monte Carlo algorithm~\cite{newman1999monte,melko2001long} (see Appendix \ref{sec: MC}); nonlocal loop updates are necessary to move between the spin configurations compatible with the constraint imposed by $K\to\infty$.
The magnetisation and heat capacity from these simulations are plotted in Fig.~\ref{fig:MCv}.
The former appears
smooth and does not saturate at any finite field [Fig.~\ref{fig:MCv}(a)], although
it does show non-analytic behaviour at a critical value of
$x_c\approx1.14$, as shown in the inset.
As the energy is entirely of Zeeman origin, criticality is also manifest as a peak in the specific heat that grows with system size [Fig.~\ref{fig:MCv}(b)] and scales {\it  with the same exponent} (see inset).  
 Remarkably, our data show the scaling is consistent with the specific heat and correlation length exponents of the 3D Ising universality class. 

This behaviour  indicates a second-order phase transition, in violation of the Landau--Ginzburg--Wilson paradigm.
Since the magnetic field selects a unique ground state, the magnetization should either smoothly approach saturation, undergo a first-order transition or accidentally pass through a critical end point to such a transition. 
We will show that it is in fact a new form of Kasteleyn mechanism~\cite{kasteleyn63a} and with the help of \textit{Kramers--Wannier duality}~\cite{Wegner1971}, we prove that it is indeed an unconventional transition giving rise to nonlocal, topological order and 3D Ising criticality.

\section{Results}%
\subsection{Kasteleyn mechanism}
A Kasteleyn transition can occur in a system with a zero-divergence constraint in an appropriate emergent field, as for example in hard-core dimer models which map to emergent U(1) gauge theories~\cite{kasteleyn63a,Jaubert2008Kasteleyn}.  In the FPM model, an analogous constraint would be the imposition of a perfect monopole crystal structure, with charge conservation on all tetrahedra. The gauge sector for these configurations, which is exactly that of hard-core dimers on a diamond lattice~\cite{brooks_artificial_2014}, orders into the fully polarised state in the presence of a  $[111]$ field  [Fig.~\ref{fig:latt}(a)]. At the transition, topological defects in the emergent field deconfine allowing for the propagation of system spanning strings of spin flips along the field direction [e.g., string I in Fig.~\ref{fig:latt}(b)]. Such an excitation costs Zeeman energy proportional to its length, the system size $L$. However, it also leads to an entropic gain in the same proportion: it can thread the kagome layer from the triangular layer above it ($\alpha=0$) in three different ways (sublattices $\alpha=1,2,3$).   
Their contribution to the free energy changes sign at the transition at $x_K=(3/8)\ln 3\approx0.412$. System-spanning strings are thus exponentially unlikely to appear at high field, resulting in perfectly saturated magnetization for field values down to $x_K$, followed by a sudden kink in the magnetization and singular behavior for the susceptibility~\cite{kasteleyn63a,SM,Nagle1975}.

\subsection{$\ztwo$ deconfinement mechanism}
This picture clearly needs to be modified for the present case (that of a liquid rather than a crystal of monopoles). For the liquid, the magnetization neither saturates at high fields nor shows a kink at the transition point (Fig.~\ref{fig:MCv}), the estimated $x_c\approx1.14$ is far from the predicted value for the Kasteleyn transition, and its Ising universality class implies isotropic (instead of \textit{anisotropic}~\cite{moessner_field_2003,baez2016}) scaling. Since $x_c>x_K$, the ordered phase is less stable, indicating additional mechanisms for thermal fluctuations.

In the FPM model~\eqref{eq:ham}, monopole charges are no longer locally conserved. 
As before, flipping an ``in-out'' pair on a tetrahedron propagates a virtual defect in the $U(1)$ field through the tetrahedron but keeps the local charge unchanged; flipping two inwards (or outwards) spins inverts its charge. This allows for \textit{any} closed loop of spins  to be flipped with loop closure ensuring global charge conservation. Starting from the fully ordered state, the lowest energy excitations are hexagon loops confined to a single kagome plane [loop III in Fig.~\ref{fig:latt}(c)] since the Zeeman energy of these spins is only a third of those in triangular layers. In addition to short loops, a system-spanning string can be excited entirely within the kagome layer [string II],  which \textit{changes} the sign of every charge it passes through. Such strings would drive a new kind of two-dimensional topological transition in an isolated layer with $x_K' = (3/2)\ln 2\approx 1.040$ (see Appendix \ref{sec: free energy argument}), a value close to the observed three-dimensional transition. On the pyrochlore lattice, these strings can jump between kagome planes by flipping a spin on an intermediate triangular layer, allowing for three-dimensional deconfinement. However, now the two classes of passage through a tetrahedron allows the deconfining virtual defect to alternate in both sign and propagation direction leading to $\ztwo$ rather than $U(1)$ deconfinement. Short loops such as types III and IV in Fig.~\ref{fig:latt}(c) re-establish the charge disorder that characterizes the monopole liquid at small $x$.
Since their energy cost is finite there is always a finite density of them ensuring that the magnetisation $M$ remains unsaturated~\cite{kasteleyn63a,bhattacharjee1983,Nagle1989,Jaubert2008Kasteleyn,smerald16a,smerald18a} for any $x$.

\subsection{Order parameters} 
\begin{figure}
    \includegraphics[width=\columnwidth]{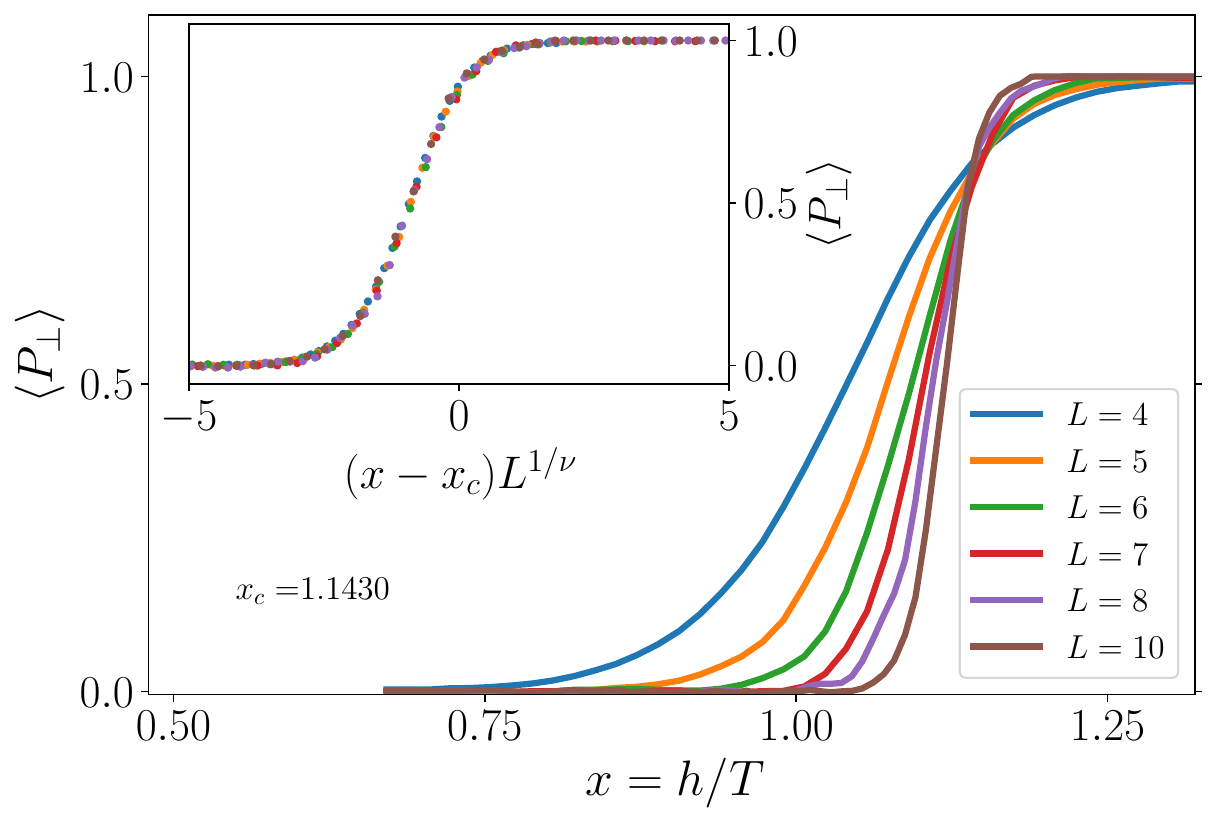}
    \caption{
    $\mathbb{Z}_2$ topological order parameter $\langle P \rangle$ for the FPM model in a $[111]$ field. 
    The inset shows the finite-size scaling with 3D Ising critical exponent $\nu$; the best visual collapse is achieved at $x_c= 1.1415$.
        }
    \label{fig:OP}
\end{figure}

The $\ztwo$ nature of the transition is highlighted by the behaviour of the \textit{parity} $P = \prod_i \sigma_i$, where the product is over all spins on a single plane of spins, kagome or triangular.  
As closed loops of finite length always flip an even number of spins in the plane, $P$ remains exactly $+1$ throughout the high field phase. At the deconfinement transition, strings that span the periodic boundaries appear for the first time: $P$ can now fluctuate between $\pm1$, and its thermal average falls to zero. This expectation is confirmed by the Monte Carlo simulation results for $\langle P_\perp \rangle$ where the chosen planes lie perpendicular to the applied field,  plotted in Fig.~\ref{fig:OP}. The inset shows a finite size scaling collapse of the data near the transition involving the correlation length critical exponent for the Ising model. Unlike the Kastelyn transition, the $\ztwo$ deconfinement transition is isotropic, so that any set of planes can be chosen.

The parameter $\langle P\rangle$ is therefore a non-local measure of the $\ztwo$ topological order, although it is not an order parameter in the thermodynamic sense. In fact, as there is no broken symmetry, emergent or otherwise, such an order parameter cannot be defined. 
In the following, we explain this unconventional transition with the help of a duality \cite{Wegner1971}.

\subsection{Kramers--Wannier duality}  
For convenience, we flip the sign convention of all spins $\sigma_i$ in the $(111)$ triangular layers ($\mathbf{e}_0\rightarrow \mathbf{e}'_0= -\mathbf{e}_0$), such that the local fields favor $\sigma'=-1$ on every site and the $K\to\infty$ constraint requires an \textit{even} number of $\sigma'=\pm1$, corresponding to the eight-vertex model on the diamond lattice. The Hamiltonian~\eqref{eq:ham} becomes
\begin{equation}
\mathcal{H}_0 = -K\sum_\mathrm{t}\sigma'_1\sigma'_2\sigma'_3\sigma'_4 - \sum_i h_{i} \sigma'_i,
\label{eq:ham2}
\end{equation}
where $h_i=h$ for spins in the triangular and $h/3$ in the kagome $(111)$ layers. Let us consider the nearest-neighbor Ising ferromagnet ($\tau_a=\pm 1$) on the diamond lattice,
\begin{equation}
\mathcal{H}^* = -\sum_{\langle ab\rangle} J_{ab} \; \tau_a \tau_b,
    \label{eq: dual Ising Hamiltonian}
\end{equation}
where the diamond sites $a,b$ are the tetrahedron centers of the original pyrochlore lattice [Fig.~\ref{fig:latt}(a)]. Following Ref.~\cite{Wegner1971}, we consider the high-temperature expansion of~\eqref{eq: dual Ising Hamiltonian} and the low-temperature expansion of~\eqref{eq:ham2}. With $N_\tau$ diamond sites, the former takes the form
\begin{equation}
  Z^*  = 2^{N_\tau} \prod_{\langle ab\rangle} \cosh(\beta J_{ab}) \Big\langle \prod_{\langle ab\rangle} [1+\tanh(\beta J_{ab}) \tau_a \tau_b] \Big\rangle, 
\end{equation}
where the expectation value is taken over the infinite-temperature ensemble. 
Upon expanding the second product, we get a term for every set of diamond-lattice bonds $\langle ab\rangle$, which contains an expectation value of the form $\langle \tau_a\tau_b\dots\rangle$.
At $T=\infty$, these expectation values vanish, unless each $\tau_a$ appears an even number of times, so that their product is identically $+1$:
That is, $Z^*$ gets contributions only from terms where each diamond-lattice site belongs to an even number of bonds, in which case these bonds form a number of closed loops. 
Therefore, 
\begin{align}
    Z^* = 2^{N_\tau} \prod_i \cosh(\beta J_i) {\sum_{\{i\}}}' \prod_i\tanh(\beta J_i),
\end{align}
where we simplify the notation from diamond-lattice bonds $\langle ab\rangle$ to the pyrochlore sites $i$ sitting on them; the primed sum $\sum'_{\{i\}}$ runs over all sets of these bonds that form closed loops.
The eight-vertex constraint ($K\rightarrow \infty$) 
imposes the same structure on the low-temperature expansion of~\eqref{eq:ham}:
\begin{equation}
Z_0 = \prod_i e^{\beta h_{i}} {\sum_{\{i\}}}' \prod_i e^{-2\beta h_{i}} \ .
\end{equation}
Since $h_{i}>0$, a Kramers--Wannier duality follows: If $\tanh (\beta J_i) = e^{-2\beta h_{i}}$, the partition functions of~\eqref{eq:ham} and~\eqref{eq: dual Ising Hamiltonian} are the same up to an overall analytic prefactor. Note however, there is no local mapping between the $\tau$ and $\sigma'$ variables as the dual  models share loop structures but not microscopic degrees of freedom.

As $J>0$ on all bonds of the diamond lattice, the Hamiltonian~\eqref{eq: dual Ising Hamiltonian} has a single ferromagnetic transition, which belongs to the 3D Ising universality class. By duality, the transition in the FPM model~\eqref{eq:ham} must therefore belong to the same universality class. The high-$h$ and low-$h$ phases are dual to the low-$J$ paramagnet and the high-$J$ ferromagnet, respectively.

The ferromagnetic order parameter of the dual model~\eqref{eq: dual Ising Hamiltonian} now becomes a non-local string order parameter in the original model:
\begin{equation}
    \langle \tau_a \tau_b\rangle_\mathrm{dual} = {Z'_{ab}}/{Z} = e^{-\beta F_{ab}},
\end{equation}
where $Z'_{ab}$ is the partition function of~\eqref{eq:ham} with two virtual defects, tetrahedra $a$ and $b$ with a vacancy or double monopole such that $\sigma'_1\sigma'_2\sigma'_3\sigma'_4 = -1$ rather than $1$.
While this might appear local, inserting such a pair of topological defects requires flipping a string of spins between $a$ and $b$.
In the paramagnet, $\langle \tau_a\tau_b\rangle$ decays exponentially with distance: that is, the free-energy cost $F_{ab}$ of the defects in the high-$h$ dual phase grows linearly, so that the virtual defects are confined.
In contrast, $\langle \tau_a\tau_b\rangle$ tends to the square of magnetic order parameter in the ferromagnet, so in the dual low-$h$ phase, their free-energy cost is finite at any separation and the defects are deconfined. 

We have tested this prediction by simulating the string correlator 
\begin{equation}
    \langle S_{ab}\rangle = \Big\langle \prod_a^b e^{-2 \beta h_i\sigma_i}\Big\rangle,
\end{equation}
for spins lying on the bonds of a simplified 8-vertex model in which $h_i=h$ on all sites and where the product runs over a fixed path between tetrahedra $a$ and $b$. The transition of this simplified 8-vertex model is in the same Universality class as the one of our model of Eq.~(\ref{eq:ham}) and is found at $x_c=h/T\approx 0.52$ (approximately half that of the FPM fluid). The finite size scaling shows an excellent collapse using the scaling exponent $2\beta/\nu$ for the 3D Ising model (see SI for details). This exponent is a function of the eigenvalue exponent in the field direction, $y_h$, showing that hidden in this non-local quantity we have the second independent critical exponent of the 3D Ising model. As a consequence, we have shown unequivocally that the FPM fluid shows a topological transition with $\ztwo$ deconfinement and that this falls in the 3D Ising universality class.

The property that the correlation functions in one model corresponds to the free energy cost of nonlocally deforming boundary conditions in its dual~\cite{Wegner1971} works in both directions.
In particular, the topological order parameter $\langle P\rangle$ is equal to the ratio of the partition function of the dual model~\eqref{eq: dual Ising Hamiltonian} with antiperiodic and periodic boundary conditions (see Appendix \ref{sec:topoOP}).
The paramagnet is insensitive to these boundary conditions, so the partition function in the thermodynamic limit is equal for both, leading to $P=+1$ in the dual high-$h$ phase. 
By contrast, ferromagnetic ordering is frustrated by the antiperiodic boundary conditions, so $\langle P\rangle$ is suppressed exponentially in the low-$h$ phase. 
Even though the system is anisotropic, the duality shows that $\langle P \rangle$ is a proper order parameter for all (111) planes, whether perpendicular to the field direction axis or not as we have shown in our numerical simulations.

\subsection{Quantum--classical mapping}%
We can also capture this transition using a field theory similar to the one developed for the Kasteleyn transition in Ref.~\cite{Powell2008KasteleynFieldTheory}, where strings of flipped spins are world-lines of hard-core bosons in (2+1)-dimensional spacetime, and the [111] field direction plays the role of imaginary time. 
However, the number of bosons (strings) is no longer conserved between imaginary-time slices: at the ends of closed loops (e.g., loop III in Fig.~\ref{fig:latt}), they can be created or removed, but only {\it in pairs}, requiring pairing terms in the effective action.
In terms of the boson field $\psi$, the most general quadratic Lagrangian thus reads
\begin{equation}
    L = \sum_k \psi_k^* \partial_\tau\psi_k + r(k) \psi^*_k \psi_k + \frac12[\Delta(k) \psi^*_k \psi^*_{-k} + \mathrm{c.c.}],
    \label{eq: Z2 Kasteleyn boson map}
\end{equation}
which can be Boguliubov transformed into
\begin{equation}
    L = \sum_k \phi^*_k \partial_\tau \phi_k + \varepsilon(k) \phi^*_k\phi_k,
    \label{eq: Z2 Kasteleyn Boguliubov}
\end{equation}
where $\varepsilon(k) = \sqrt{r^2(k) - \Delta^2(k)}$. 
For a gapped dispersion of $\varepsilon(k)$,  finite loops may appear due to the pairing term, but none of these span the system so that the topological order parameter $P=+1$.
The transition corresponds to $\varepsilon(k=0)\to 0$: $\phi_{k=0}$ condenses, and system-spanning loops appear.
For generic dispersions $r(k), \Delta(k)$, $\varepsilon(k)\propto|k|$ at the transition, so the dynamical exponent of the transition is $z=1$, rather than the Kasteleyn value $z=2$~\cite{Powell2008KasteleynFieldTheory}.
More importantly, the action~\eqref{eq: Z2 Kasteleyn Boguliubov} and the Boguliubov transformation from $\psi$ to $\phi$ becomes nonanalytic, thus $\langle\phi\rangle$ becomes a hidden, nonlocal order parameter.

\section{Discussion}%
We have shown how a magnetic field can induce a $\mathbb{Z}_2$ topological phase transition in a variant of spin ice.  The transition is reminiscent of that found in the toric code model with an applied field~\cite{trebst2007breakdown}: while it is detectable through a divergence in the heat capacity, there is no local order parameter. In our case, we proved that the change is driven by a very unusual Kasteleyn mechanism that is isotropic and allows excitations on both sides of the transition. Linking Kasteleyn's and Wegner's work, we demonstrated these properties through a Kramers--Wannier duality with the diamond-lattice Ising ferromagnet. This duality shows that the topological transition belongs to the 3D Ising universality class. 

Remarkably, a straightforward measurement of the magnetization in our model would find standard trappings of critical phenomena with scaling behavior, etc. However, it would determine the critical exponent for the susceptibility to be that of the specific heat, $\alpha\approx0.110$, rather than the susceptibility exponent of the Ising ferromagnet, $\gamma\approx1.237$.

Finally, we remark that extensive research on frustrated systems over many years has brought to light a variety of mechanisms to stabilize  gauge-charge excitations at low temperatures -- magnetic field \cite{matsuhira_field_2002}, further neighbor exchange \cite{udagawa_interactions_2016,rau_interactions_2016,borzi2016intermediate}, magneto-elastic coupling \cite{khomskii12a,jaubert2015,slobinsky2021,slobinsky2019_polarized}, material design \cite{lefrancois17a,miao2020two,nutakki23b,Pearce2022}, etc.
This, as well as progress in the construction of constrained models on versatile AMO~\cite{qscars_expt,ebadi_expt} and NISQ~\cite{nisq_kitaev_expt,chamon_expt} platforms, makes us optimistic about future experimental realization of our model. The present work is just one example of the rich physics that can arise in systems where emergent gauge charges are fundamental building blocks rather than rare excitations.

\begin{acknowledgements}
We thank Fr\'ed\'eric Mila for useful comments. 
A.Sz. was supported by Ambizione grant No. 215979 by the Swiss National Science Foundation.    
L.D.C.J.\ acknowledges financial support from ANR- 18-CE30-0011-01 and ANR-23-CE30-0038-01 and thanks IFLYSIB (CONICET-UNLP) for their hospitality during this project. This work was in part supported by the Deutsche Forschungsgemeinschaft under grants FOR 5522 (project-id 499180199) and the cluster of excellence ct.qmat (EXC 2147, project-id 390858490). PCWH acknowledges ANR grant No. ANR-19-CE30-0040 and grant NSF PHY-2309135 to the Kavli Institute for Theoretical Physics (KITP).
\end{acknowledgements}
\bibliographystyle{apsrev4-2}
\bibliography{biblio}

\appendix
\section{Details of the Kramers--Wannier duality} \label{App_A}

In the main text, we showed that the partition functions of the eight-vertex model
\begin{equation}
\beta\mathcal{H} = -K\sum_\mathrm{t}\sigma_1\sigma_2 \sigma_3\sigma_4 - \sum_i h_i \sigma_i\quad (K\to+\infty)
\label{eq:ham2}
\end{equation}
and the diamond-lattice Ising ferromagnet
\begin{equation}
\beta\mathcal{H}^* = -\sum_{\langle ab\rangle} J_{ab} \; \tau_a \tau_b
    \label{eq: FM Hamiltonian}
\end{equation}
are
\begin{subequations}
\begin{align}
    Z &= \prod_i e^{h_i} {\sum_{\{i\}}}' \prod_i e^{-2 h_i} ,
    \label{eq: LTE Z}\\
    Z^* &= 2^{N_\tau} \prod_i \cosh J_i  {\sum_{\{i\}}}' \prod_i\tanh J_i 
    \label{eq: HTE Z}
\end{align}
\label{eq: Z}%
\end{subequations}
in low- and high-temperature expansion, respectively. The sums are taken over all possible sets ${\{i\}}$ of pyrochlore sites $i$ and, equivalently, on diamond-site bonds in \eqref{eq: FM Hamiltonian}, with an even number of them on each tetrahedron or diamond site; in other words, over all possible sets ${\{i\}}$ of closed loops. For brevity, in this section we incorporate the inverse temperature $\beta$ into the definition of the couplings $h,J$ and drop the primes from $\sigma'$.

The Kramers-Wannier duality follows if we set
\begin{equation}
    e^{-2 h_i} = \tanh  J_i,
    \label{eq: Kramers Wannier}
\end{equation}
so that the sums in~\eqref{eq: LTE Z} and \eqref{eq: HTE Z} become identical.
The prefactors  yield a contribution to the free energy that is smooth in $h$ and $J$, so the universal properties of the two models are the same. 
In particular, let us define the modified partition functions~\cite{Wegner1971}
\begin{subequations}
\begin{align}
    Y &=  \prod_i \cosh(2h_i)^{-1/2} \times Z  = \prod_i \frac{e^{h_i}}{\sqrt{\cosh(2h_i)}} \times \Sigma\\
    Y^* &=  \prod_{\langle ab\rangle} \cosh(2J_{ab})^{-1/2} \times Z^* = \prod_{\langle ab\rangle} \frac{\sqrt2 \cosh J_{ab}}{\sqrt{\cosh(2J_{ab})}} \times\Sigma^*,
\end{align}
\label{eq: Y}%
\end{subequations}
where $\Sigma, \Sigma^*$ stand for the sum over $\{i\}'$ in~\eqref{eq: Z}. Now, if $h_i$ and $J_{ab}$ satisfy the duality condition~\eqref{eq: Kramers Wannier}, $\Sigma=\Sigma^*$ and
\begin{align*}
    \frac{e^h}{\sqrt{\cosh(2h)}} &= e^h \sqrt{\frac{2}{e^{2h}+e^{-2h}}} = \sqrt{\frac{2}{1+e^{-4h}}} = \sqrt{\frac{2}{1+\tanh^2 J}} \\ &= \sqrt{\frac{2\cosh^2 J}{\cosh^2 J+\sinh^2 J}} = \sqrt2 \frac{\cosh J}{\sqrt{\cosh(2J)}},
\end{align*}
so $Y=Y^*$. This completes the duality mapping between $\mathcal{H}$ and $\mathcal{H}^*$.

\subsection{Correlation functions}

For finite systems, the partition functions~\eqref{eq: Z} are analytic in the couplings $h,J$, so the duality $Y=Y^*$ holds for all values that obey~\eqref{eq: Kramers Wannier}, including complex ones. Following Ref.~\cite{Wegner1971}, we use this fact to compute correlation functions in our models.

Consider a deformed version $\mathcal{H}^{*\prime}$ of~\eqref{eq: FM Hamiltonian}, where the signs of some $J$ are flipped. The ratio between the partition functions of these two models is
\begin{equation}
    \frac{Y^{*\prime}}{Y^*} = \frac{Z^{*\prime}}{Z^*} = \Big\langle \prod_{\langle ab\rangle'} e^{-2 J_{ab}\tau_a\tau_b} \Big\rangle,
    \label{eq: deformation correlator}
\end{equation}
where the product is over the $N'$ terms with flipped signs. The duality relation~\eqref{eq: Kramers Wannier} maps the flipped $J$ to
\begin{equation}
    \tanh (-J) = -\tanh J = -e^{-2h} = e^{-2h-i\pi} \implies h'= h+i\pi/2,
\end{equation}
so we have
\begin{equation}
    \frac{Y'}{Y} = (-i)^{N'} \frac{Z'}{Z} = (-i)^{N'} \Big\langle \prod_{i} e^{i\pi \sigma_i/2} \Big\rangle = \Big\langle \prod_{i'} \sigma_i \Big\rangle
    \label{eq: spin correlator}
\end{equation}
as $(-i)e^{\pm i\pi/2} = (-i)(\pm i) = \pm 1$.
Comparing~(\ref{eq: deformation correlator},~\ref{eq: spin correlator}) yields
\begin{subequations}
\begin{equation}
    \frac{Z^{*\prime}}{Z^*} = \Big\langle \prod_{\langle ab\rangle'} e^{-2 J_{ab}\tau_a\tau_b} \Big\rangle = \Big\langle \prod_{i'} \sigma_i \Big\rangle.
    \label{eq: correlator duality 1}
\end{equation}
Finally, we note that~\eqref{eq: Kramers Wannier} can be rearranged to $\tanh h_i = e^{-2J_{ab}}$, which allows us to run the same argument in reverse to get
\begin{equation}
    \frac{Z'}Z = \Big\langle \prod_{i'} e^{-2 h_i \sigma_i} \Big\rangle = \Big\langle \prod_{\langle ab\rangle'} \tau_a \tau_b\Big\rangle,
    \label{eq: correlator duality 2}
\end{equation}
\end{subequations}
where $Z'$ and $Z$ differ in flipping the sign of $h_i$ on the primed pyrochlore sites.

\subsection{Topological order parameter}
\label{sec:topoOP}

Let us first apply this duality to the topological order parameter $P= \prod_i\sigma_i$, where the product runs over a \textit{reference plane} that cuts through one of the periodic boundary conditions. 
Eq.~\eqref{eq: correlator duality 1} implies $\langle P\rangle = Z^{*\prime}/Z^*$, where $H^{*\prime}$ and $H^*$ differ in flipping the sign of $J$ across the diamond-lattice bonds that cut through the reference plane. 
We can undo this change in the Hamiltonian at the cost of flipping the sign of every spin $\tau$ above the reference plane, so that the sign of $\tau_a\tau_b$ flips for precisely those bonds that cut through it.
Doing so on a torus turns the periodic boundary condition across the reference plane into an antiperiodic one:
that is, $\langle P\rangle$ is the ratio of the dual partition function in antiperiodic and periodic boundary conditions, as asserted in the main text.

We also argued in the main text that in the thermodynamic limit, $\langle P\rangle$ jumps from $+1$ to 0 as system-spanning strings proliferate at the Kasteleyn transition.
We can also understand this from the point of view of the dual Ising model.
Let the transfer matrix of one (translationally repeated) layer of the diamond lattice be $T$, and let $X$ be an operator that flips $\tau$ on every diamond site.
Then, the partition function of an $n$-layer diamond lattice is $(T^n)$ in periodic and $(T^n X)$ in antiperiodic boundary conditions. 
Furthermore, $X$ and $T$ commute due to Ising symmetry ($XTX$ is the transfer matrix of a model where all the $\tau$ are flipped, which is identical to the original model), so every eigenvector of $T$ is also an eigenvector of $X$; since $X^2 = 1$, the possible eigenvalues are $\pm1$.
\begin{itemize}
    \item In the paramagnetic phase, $T$ has a gapped spectrum, with a unique leading eigenvalue $\lambda$ in the thermodynamic limit. Therefore, $Z^* \to \lambda^n$ and $Z^{*\prime} \to x\lambda^n$ in the thermodynamic limit, where $x$ is the corresponding eigenvalue of $X$. Since $Z^{*\prime}$ is a well-defined, positive partition function, $x$ must be $+1$, so $\langle P\rangle\to +1$.
    Any subleading correction is suppressed as $(|\lambda_2|/\lambda)^n$, where $\lambda_2$ is the next-largest eigenvalue of $T$:
    This is consistent with the exponential suppression of system-spanning strings due to their linearly growing free-energy cost.
    \item In the ferromagnetic phase, spontaneous symmetry breaking requires a degeneracy in the spectrum: two eigenvectors, with $x=\pm1$, must have the same leading eigenvalue $\lambda$ in the thermodynamic limit. As a result, the leading $\lambda^n$ contributions to $Z^*$ cancel from $Z^{*\prime}$, so $\langle P\rangle\to 0$.
\end{itemize}
At the transition, the scale invariance of the problem leads to a distribution of large eigenvalues of the transfer matrix that results in a system-size-independent value of $\langle P\rangle$~\cite{Hasenbusch1993BoundaryConditions}. As a result, we expect a finite-size scaling form similar to that of the Binder cumulant:
\begin{equation}
    \langle P\rangle(h, L) \sim f_P[(h-h_c)L^{1/\nu}].
    \label{eq: topo OP scaling form}
\end{equation}

\subsection{String order parameter}

We can express any two-point correlator $\langle \tau_a\tau_b\rangle$ in the dual Ising model as a product of nearest-neighbour two-point functions along a path connecting $a$ and $b$. Applying~\eqref{eq: correlator duality 2} to this expectation value gives
\begin{equation}
    \langle S_{ab}\rangle = \langle \tau_a \tau_b\rangle = \langle (\tau_a\tau_c)(\tau_c\tau_d)\cdots(\tau_e\tau_b) \rangle = \frac{Z'_{acd\dots eb}}{Z},
    \label{eq: string OP expectation form}
\end{equation}
where $acd\dots eb$ is a path of nearest neighbours on the diamond lattice between $a$ and $b$, and the Hamiltonian for $Z'_{acd\dots eb}$ differs from $\mathcal{H}$ by flipping the sign of $h_i$ for the pyrochlore sites on the bonds $ac,cd,\dots,eb$.
Similar to the case of $\langle P\rangle$, we can eliminate this deformation in the Hamiltonian by flipping $\sigma$ on these sites.
Along the path, this does not affect the $\ztwo$ constraint (e.g., the parity of tetrahedron $c$ is unchanged if we flip $\sigma$ on pyrochlore sites $ac$ and $cd$). 
At the endpoints of the string, by contrast, we only flip one spin, which introduces a $\ztwo$ defect.
That is, $Z'$ is the partition function of $\mathcal H$ with two defects on tetrahedra $a$ and $b$, justifying the form given in the main text.
We also note that $Z'$ only depends on the positions of the defects, justifying the simplified notation $Z'_{ab}$.

$\langle S\rangle$ is really an order parameter in the thermodynamic, infinitely distant defect limit, where $\langle S\rangle\sim M^2\sim (h_c-h)^{2\beta}$, as $M$ is the spontaneous magnetisation of the dual Ising model.
On a finite system, we can define a surrogate order parameter $\langle S\rangle_L$ by choosing defect sites at the largest possible distance. For these, we expect the finite-size scaling form of the squared magnetisation to apply:
\begin{equation}
    \langle S\rangle_L(h) \sim L^{-2\beta/\nu} f_S[(h-h_c)L^{1/\nu}].
\label{eq: string OP scaling form}%
\end{equation}


\section{Estimates of the critical value of the tuning variable}
\label{sec: free energy argument}

\begin{figure}
\includegraphics[width=\columnwidth]{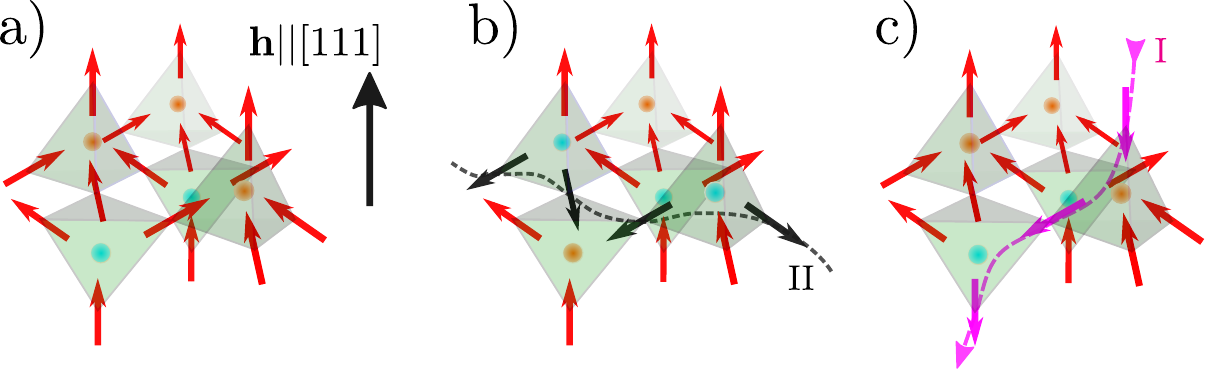}
\caption{a) Initial state, saturated by a $[111]$ magnetic field $\mathbf{h}$ with single positive (orange) and negative (cyan) gauge charges forming a zincblende structure. The magnetic field is parallel to the apical spins in triangular planes, with basal spins forming kagome lattices.
b) On decreasing $h/T$, the system demagnetizes, constrained to retain a single gauge charge in each tetrahedron. Here, we propose a ``kagome string'' (marked II, in black) contained entirely in one of the aforementioned kagome planes. This type of excitation leads to an inversion of the topological charge in each tetrahedron it goes through.
c) A ``[111] string'' (marked I, in magenta) runs parallel to $\mathbf{h}$; it is composed of alternating apical and basal spins, and leaves the charge distribution unchanged. However, flipping such a string requires more energy due to the higher Zeeman coupling of apical spins.
}
\label{fig:string}
\end{figure}

As mentioned in the main text, the system has a single tuning variable $x=h/T$. A simple estimate for its critical value ($x_c$) can be made if we assume that
\begin{enumerate}
    \item the system is fully polarized for $x>x_c$ [Fig.~\ref{fig:string}(a)];
    \item on lowering $x$, the first string excitation propagates entirely within a kagome plane perpendicular to [111] (a \textit{kagome string}), so that it is made entirely of energetically cheap basal spin flips [Fig.~\ref{fig:string}(b)].
\end{enumerate}
Thinking of the process of string formation as a succession of these basal spin flips~\cite{Jaubert2008Kasteleyn}, we notice that each step implies choosing between two equivalent spins within a given triangle of the kagome plane involved. The associated entropy change is then $\delta s=\ln(2)$. On the other hand, the Zeeman energy cost of flipping the spin is $\delta u = 2h/3$. The total free energy change coming from introducing a closed string of $l$ spins on a kagome plane is thus
\begin{equation}
\delta \mathcal{F}= l\, \delta \!f = l \, (2h/3 - T \ln(2)).
\label{eq:deltaf}
\end{equation}
In the thermodynamic limit, the length of a loop winding across the periodic boundary conditions diverges, $l\to\infty$, so a positive free-energy cost $\delta \!f > 0$ prevents them from appearing. On the other hand, a negative $\delta \!f < 0$ favors their creation; their proliferation is only limited by entropic interactions between close-by strings. The Kasteleyn transition occurs at $\delta\!f=0$, that is, at $x_c = h_c/T = 3/2\ln(2) \approx 1.04$, as stated in the main text. 

Note that, different from the usual Kasteleyn transition, Eq.~\eqref{eq:deltaf} can also be applied to any closed loop [e.g., loop III in Fig.~1(c) in the main text]; since they only cost a finite amount of free energy ($l$ can now be finite, even in the thermodynamic limit), they are present at any $x$, even above the transition point. In light of this, we understand that our first assumption is just a convenient approximation, in spite of which we obtain a value of $x_c$ within $10\%$ of the measured value.

A similar reasoning can also be used to estimate the critical coupling for introducing strings of flipped spins running along the field direction (\textit{[111] strings}). 
We still use assumption~1; the kagome string described in assumption~2 is replaced with one propagating along $[111]$, made of alternating energetically expensive apical spins (with \textit{no} entropy gain) and cheap basal spin flips (with three possible exit ways) [Fig.~\ref{fig:string}(c)].
The  entropy gain and energy cost per two consecutive flips are $\delta s=\ln(3)$ and $\delta u = (2+2/3)h=8h/3$, respectively. 
That is, we would expect a transition at $x\approx x_K = (3/8)\ln 3\approx0.412$ based on [111] strings alone. If this were the main mechanism, the transition would only occur at $h/T$ less than half the value we actually observed.

Our next task is to understand why the estimation for $x_c$ for kagome strings gives a reasonable number, while that for [111] strings ($x_K$) is so bad. This will also provide a hint of why both [111] and kagome strings start winding around the system at the same value of $x$ in the thermodynamic limit.  The missing ingredient is that even strings winding along [111] could meander along perpendicular kagome planes. Below $x_c$, these kagome-string segments give rise to a negative free-energy contribution: 
If they are long enough, they can compensate for the free-energy cost of flipping the apical spins, thus facilitating system-spanning loops in all directions.


\section{Details of the Monte Carlo simulations}

\label{sec: MC}
Since the four-body interaction $K\rightarrow \infty$ forbids all single-spin flips, the data shown in this paper were obtained from Monte Carlo simulations based on a loop algorithm~\cite{newman1999monte,melko2001long}. In order to construct a loop, we start a random walk on an arbitrary initial spin $\sigma_i$ (spin number $s=1$) from fully-packed monopole configuration. The second spin in the chain ($s=2$) is chosen at random out of its six nearest neighbors. Different to the spin ice case, the single-charge constraint is preserved for \textit{any} of these spins, either preserving the local monopole charge (if one spin points in and the other out of the common tetrahedron) or inverting it (if both pseudospins $\sigma$ have the same sign). 
From here on, spin $s+1$ is chosen at random out of the three spins that share a tetrahedron with spin $s$ but not with $s-1$.
The process terminates when a spin $s_l$ is visited twice. All spins with $s<s_l$ are then dropped, leaving a closed loop between the two occurrences of spin $s_l$. In other words, this is a \textit{short-loop} algorithm~\cite{newman1999monte}. These loops can be as short as 6 spins [Fig. 1(c) in main text], or can wind the system along any direction through PBC, one or more times. A new monopole-liquid configuration is then proposed by flipping all spins in the loop and accepted or rejected according to the Metropolis rule, with energy changes arising only from the Zeeman contribution.

\begin{figure}
    \includegraphics[width=\columnwidth]{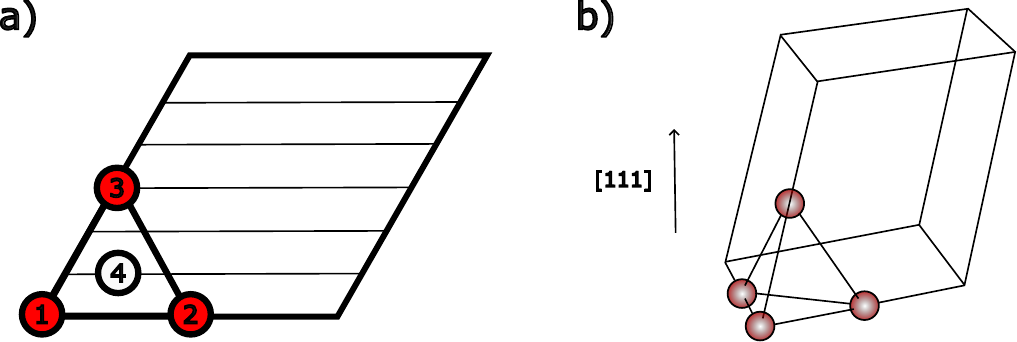}
    \caption{The non-orthogonal (rhombohedral) unit cell, with 4 spins: a) 2D projection down a $\langle 111\rangle$ direction; b) three-dimensional view.
    We used simulation boxes of rhombohedral shape, made up of $L$ unit cells in each direction, with periodic boundary conditions.}
    \label{fig:NonOrth-UC}
\end{figure}

The data presented in the main text were collected using simulation boxes consisting of $L\times L\times L$ of the rhombohedral unit cells shown in Fig.~\ref{fig:NonOrth-UC}, with periodic boundary conditions (PBC) across all edge directions of the rhombohedron. Each curve is an average of at least 10 runs, with a minimum of $1 \times 10^6$ steps each, where one step is defined as $N$ attempts to introduce a loop to the system, with $N$ the number of sites in the lattice.

A different set of measurements were performed with the conserved monopoles algorithm~\cite{borzi_2013}, using the conventional cubic unit cell and PBC along $\langle 100 \rangle$ directions. Here, dynamics is restricted in a way such that two neutral charges are allowed to wander among a liquid of magnetic monopoles, preserving the number of magnetic charges. In the limit of large $L$ (in which the two neutral sites make a negligible contribution to any thermodynamic quantity) we obtained identical results to those the loop algorithm. An exception is the topological order parameter $\langle P\rangle$: due to the two neutral sites, $P$ would be different on different $(111)$ planes, making the order parameter ill-defined.

\subsection{Finite-size scaling of thermodynamic data}

In Fig.~2 of the main text, we performed finite-size scaling collapse on the measured heat capacity and magnetisation data. Since it is not an order parameter of the transition, the heat capacity follows the scaling form
\begin{subequations}
\begin{align}
    C(x=h/T) &= C_0(x) + A_\lessgtr |x-x_c|^{-\alpha}
    \\
    C(x, L) &= C_0(x) + L^{\alpha/\nu} f_C[(x-x_c)L^{1/\nu}]
\end{align}
\end{subequations}
in the thermodynamic limit and for finite systems, respectively.
Here, $C_0(x)$ is a nonsingular contribution without significant finite-size scaling, also expected to be nonzero far from the transition.
Usually, we can perform a finite-size scaling analysis by plotting $CL^{-\alpha/\nu}$ as a function of $(x-x_c)L^{1/\nu}$, effectively neglecting the nonsingular $C_0$.
For the 3D Ising transition, however, $\alpha=0.110$, so the singular peak grows very slowly for numerically practical system sizes:
in the presence of $C_0$, it is difficult to achieve tolerable scaling collapse to any critical exponent, or to the log-scaling typical at $\alpha=0$.
To eliminate this correction, it is natural to subtract $C(x,L)$ measured at different system sizes; to recover the familiar scaling form, we need to keep the \textit{ratio} of system sizes fixed:
\begin{align}
    C(x,L) - C(x,qL) = L^{\alpha/\nu} f_C[(x-x_c)L^{1/\nu}] \notag \\ 
    - q^{\alpha/\nu}L^{\alpha/\nu} f_C[(x-x_c)q^{1/\nu}L^{1/\nu}] = \notag \\ L^{\alpha/\nu} \tilde f_C[(x-x_c)L^{1/\nu}],
    \label{eq: fss C}
\end{align}
where
\begin{equation*}
    \tilde f_C(y) = f_C(y) - q^{\alpha/\nu} f_C(yq^{1/\nu}).
\end{equation*}
In Fig.~2, we fixed the ratio $q=1/2$; as shown in the inset, the heat capacity differences fit well to the scaling form~\eqref{eq: fss C}. 
We used the same technique to perform finite-size scaling on the magnetisation as well. Here, the relevant critical exponent\footnote{Note that in our model, the magnetisation is not an independent order parameter, but simply a rescaling of the total energy.} is $\alpha-1<0$, so the singular component becomes smaller as the system size increases. Therefore, the subtraction scheme is needed to eliminate the nonsingular component that dominates the thermodynamic limit.


\section{Simulations of the isotropic eight-vertex model}

In addition to the model discussed in the main text, we also performed simulations of the isotropic eight-vertex model $h_i\equiv h$.
This model is dual to the isotropic diamond-lattice Ising model, for which the transition point $J_c=0.36973980(9)$ is accurately known~\cite{Deng2003CriticalPoint}. This implies a transition at $h_c = 0.51956249(11)$;%
\footnote{We note that a variant of the arguments in Sec.~\ref{sec: free energy argument} applies here too. Now, every flipped spin costs energy $2h$, and after entering a tetrahedron, the string excitation can exit it equally well through all three remaining spins. Therefore, the free-energy cost of a string of length $l$ is approximately $l(2h-T\ln 3)$, which implies a transition at $h/T = \ln3/2 \approx 0.549$, within 10\% of the true value.}
together with the known Ising critical exponents, this allows us to test the predicted scaling forms \textit{without any fitting parameters.}

We performed simulations on cubic clusters of $L\times L\times L$ conventional unit cells of the pyrochlore lattice ($16L^3$ spins) using a variant of the loop algorithm of Sec.~\ref{sec: MC}. We decompose the system into a dense covering of closed loops by connecting two pairs of spins on every tetrahedron; these loops are then all proposed to be flipped according to the Metropolis rule. To help equilibrate the topological parity $P$, we also propose a number of randomly selected minimal system-spanning loops in each Monte Carlo step.
In addition, we found that running every temperature point simultaneously in a parallel tempering scheme greatly improves mixing in the high-$h$ phase, where successfully inserting or removing a system-spanning loop becomes exponentially unlikely.

\begin{figure}
    \centering
    \includegraphics{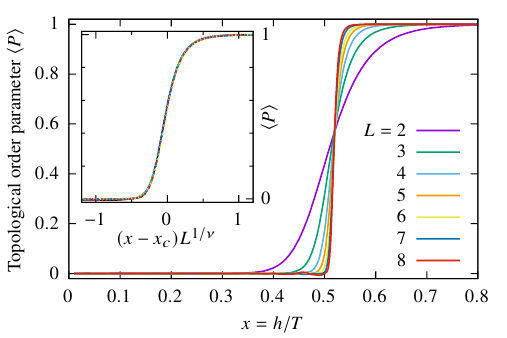}
    \includegraphics{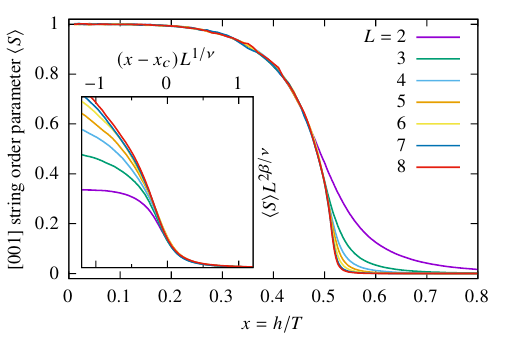}
    \caption{
    Left: The $\mathbb{Z}_2$ topological order parameter $\langle P \rangle$ from simulations of the isotropic eight-vertex model in cubic clusters of $L\times L\times L$ unit cells ($16L^3$ spins).
    Right: The string order parameter~\eqref{eq: string OP expectation form} for a pair of sites separated by $L/2\times [001]$ (i.e., halfway across the periodic boundary conditions). 
    Insets: finite-size scaling collapse with 3D Ising critical exponents $\beta,\nu$ and the critical field $x_c = 0.51956249(11)$ predicted by Kramers--Wannier duality.
    }
    \label{fig: isotropic}
\end{figure}

In the cubic simulation box, the topological order parameter $\langle P\rangle$ is defined for system-spanning (100) planes, all of which are equivalent; the measured results are shown in Fig.~\ref{fig: isotropic}.
Similar to the anisotropic case, $\langle P\rangle$ tends to 1 and 0 above and below the transition, respectively, with an increasingly sharp jump between the two as the system size is increased. Indeed, the data fit the predicted scaling form~\eqref{eq: topo OP scaling form} perfectly.

We also measured the string order parameter~\eqref{eq: string OP expectation form}. Naïvely, one could measure $Z'_{ab}/Z$ by pairing up the spin configurations that contribute to the two partition functions, which leads to the Monte Carlo estimator~[cf.~\eqref{eq: correlator duality 2}]
\begin{equation}
    \langle S_{ab}\rangle = \Big\langle \prod_a^b e^{-2 h_i\sigma_i}\Big\rangle,
\end{equation}
where the product runs over a fixed path between tetrahedra $a$ and $b$.
However, in the deconfined phase, the total spin $\sum_i\sigma_i$ might take a range of positive and negative values: after exponentiation, this leads to a huge variance in the estimator, which makes it extremely impractical to use.
Instead, we introduced an additional update to our Monte Carlo algorithm, which allows flipping a fixed, open string, thereby inserting or removing a pair or defects at its ends. This way, we sample the partition function $Z'_{ab}+Z$, so the ratio $Z_{ab}/Z$ can be estimated as the ratio of samples with and without defects. This estimator (combined with parallel tempering) converges quickly and reliably to the true value of $\langle S\rangle$.

To obtain an estimate of the long-range string order parameter, we measured $\langle S\rangle$ for two defects separated by $L/2\times[001]$, that is, halfway across the periodic boundary conditions.%
\footnote{For odd $L$, there are no pyrochlore sites separated by precisely this distance. Instead, we have used one of the six sites nearest to the halfway point. We cannot see any resulting even-odd effect in the data.}
These are plotted in Fig.~\ref{fig: isotropic}: In the high-$h$ phase, they tend rapidly to zero, while below $h_c$, they converge to a finite value, which appears to scale as a power of $(h_c-h)$ near the transition. 
Apart from saturation in smaller systems, the data also obey the finite-size scaling form~\eqref{eq: string OP scaling form}.

\end{document}